# FPGA implementation of a DCDS processor

Simon Tulloch
European Southern Observatory, Karl Schwarzschild Strasse 2, Garching, 85748, Germany.

**Abstract.** An experimental digital correlated double sampler (DCDS) video processor has been implemented in a Xilinx Artix FPGA. It uses an Opal Kelly XEM7010-A50 module that comes with an integrated USB2 interface for easy interfacing to a data acquisition PC. The FPGA has been coupled to a dual 16-bit 20Msps ADC and video preamplifiers. A synthetic CCD video waveform was provided by a high speed DAC also under FPGA control. VHDL-defined hardware implemented the differential averager algorithm which has been shown to give the optimal signal to noise ratio over a wide range of readout speeds. Hardware developed in this project is intended for eventual integration into a Xilinx Zynq-based system that combines the functionality of both a CCD controller and Linux-based data acquisition system.

**Keywords:** DCDS, CCD, FPGA.

## 1    The development board

The processor implementation was developed on the hardware shown below in figure 1. This consists of a baseboard containing two video preamplifiers, a 2-channel 16-bit 20Msps ADC, a 2-channel 8-bit DAC and a switch mode power supply. Piggy-backed on this board is an Opal Kelly XEM7010 FPGA module with a USB interface to the host PC. The FPGA is an xc7a50tifgg484-1L, which is at the low end of the Artix-7 range but still usable with the Vivado design suite. The FPGA module came with a very useful interface package known as "FrontPanel" to ease data transfer to the host PC.

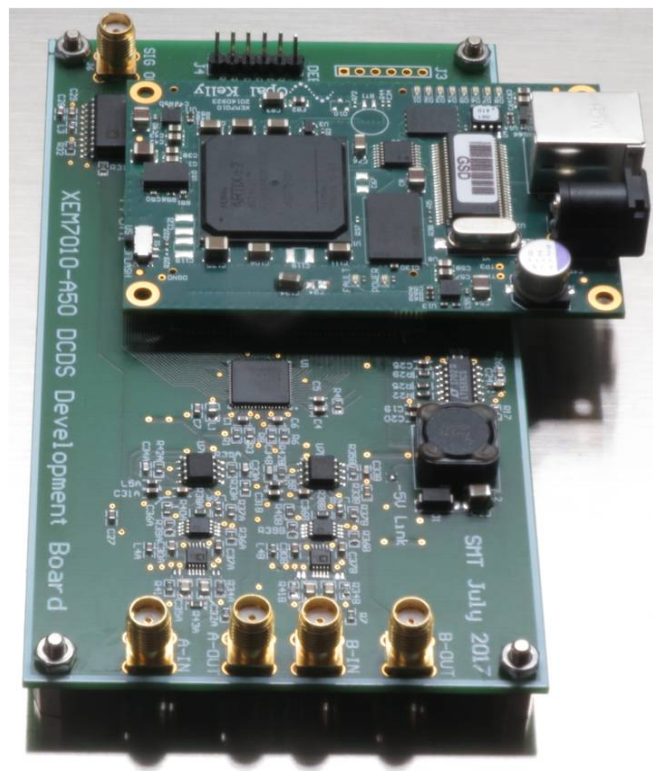

**Fig. 1.** The development board.

## 2   Video Preamplifier Design

The development board contains two fully-differential video preamplifiers whose design is shown in figure 2. The front end uses a dual-package low-noise op-amp of which several pin compatible variants were experimented with. The gain was calculated to give a full range signal at the ADC in response to a 830mV input. The bandwidth was set to 4MHz, well below the Nyquist frequency of the ADC. A baseline restoration switch was included right at the front end to periodically replenish the charge on the input coupling capacitors that is lost to op-amp input-bias current. The front end op-amp is supplied from split rails for optimum performance. The second stage ADC driver, which could equally well function as a cable driver, uses a single rail supply so as to avoid damaging the ADC by the application of a negative voltage to its analogue inputs. This circuit owes much to previously published designs[1].

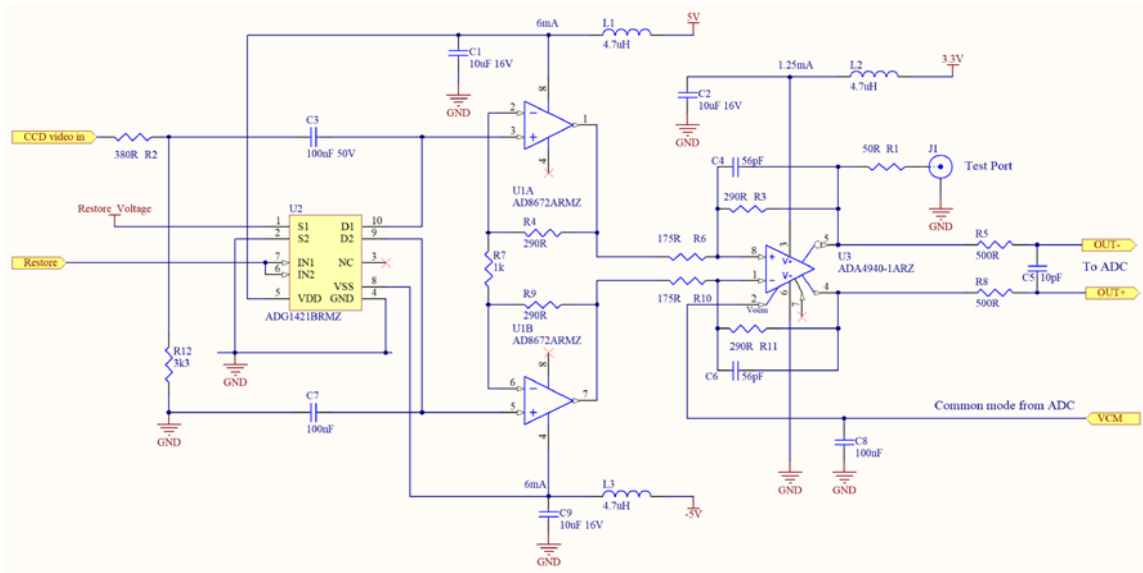

**Fig. 2.** The pre-amplifier circuit

### 2.1   Need for a black clamp

The baseline restoration circuit can also function as a black clamp to temporarily "blind" the amplifier when the CCD is being vertically clocked and the video waveform heavily contaminated with clock feed-through. This preserves the quality of the first few image columns. In preamps with lower supply rails the black clamp has the additional advantage of potentially doubling the dynamic range of the system.

### 2.2   Noise spectra of preamplifiers

This was measured by first transmitting raw ADC values to the host PC, effectively turning the development board into a high resolution oscilloscope. A Python program was then used to FFT the waveform to produce a properly scaled noise spectrum. Five hundred spectra were then averaged to reduce measurement noise. The poor performance of the supposedly low-noise switch mode on-board inverting regulator (LT1534) was immediately apparent and this had to be replaced with a bench power supply in order to faithfully measure the intrinsic preamplifier noise. Spectra were obtained with both the AD8056 and the AD8672 and at differential gains of x2.4, x4.9 and x9.6 corresponding to a system sensitivity of 1.7, 0.8 and 0.4$e^-$/ADU respectively (assuming CCD node sensitivities of 7.5 $\mu$V/$e^-$). These were plotted together with the known noise spectral density of the E2V CCD231 for comparison. This is shown in figure 3.  The input referenced noise decreased as the differential gain increased suggesting that the limiting noise came not from the preamp but the ADC driver or the ADC itself. A differential gain of x4.9 nevertheless placed the system noise well below the noise spectrum of a typical low-noise CCD. Given that the differential input range of the ADC used (AD9269) was 2V this restricted the dynamic range to about 54k photo-electrons which is too low. It may have been better to use an ADC with a larger differential input range to permit higher front-end gains. The Analog Devices AD7626 would be one option although maximum conversion rate of this single channel device is only 10MHz.

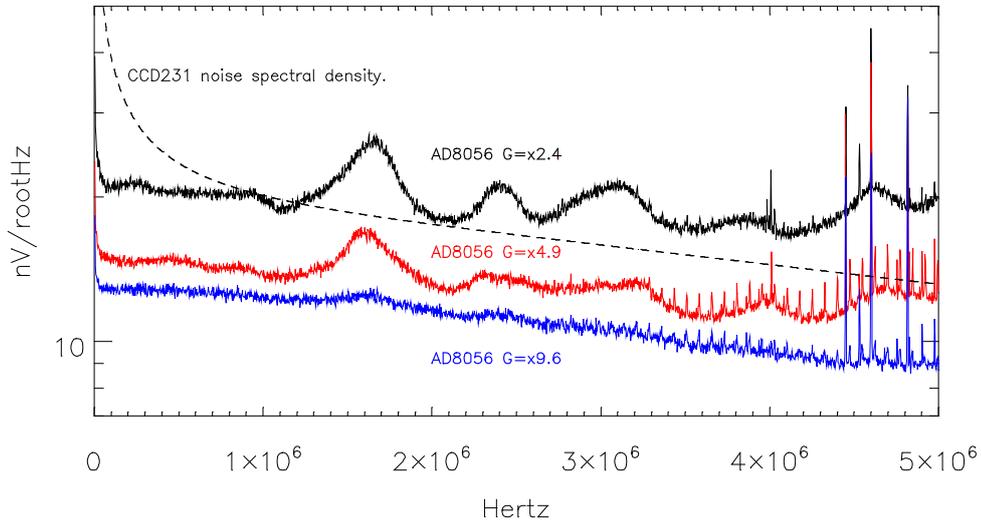

**Fig. 3.** Noise spectra of the preamp at various differential gains.

## 2.3 Bandwidth of preamplifiers

This was measured by injecting a swept frequency sine wave at the input and measuring its output amplitude using an oscilloscope. The baseline restoration circuit was disabled during these tests and the op-amp input bias current supplied through 100k resistors to ground.

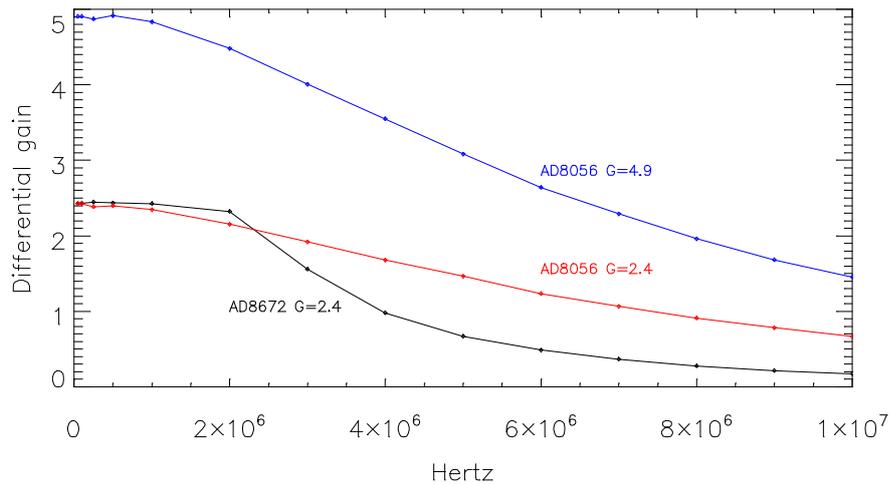

**Fig. 4.** Bandwidth of preamp variants.

The AD8056 , with its 300MHz gain-bandwidth product, saw little change in bandwidth in both the x2.4 and x4.9 gain configurations. The slower AD8672 had a bandwidth of only 2.5MHz even with the lower gain setting  of x2.4, limited by its intrinsically low gain-bandwidth product.

## 3 DCDS Algorithm

The algorithm provides flexibility on how the ADC samples in each pedestal are weighted. Up to 256 samples can be taken from each pedestal and each sample can be given a weighting of between 0 and 255. The sample weights are downloaded from the host upon power-up. One important feature is that regardless of the weight values, the gain of the DCDS processor will be x1. This is accomplished by calculating the reciprocal of the sum of the coefficients as soon as they are downloaded. This reciprocal is then multiplied by the processor output prior to each pixel's transmission in order to normalise its gain. Calculating the reciprocal is quite intensive requiring a pipeline of 66 cycles, however , it only needs to be done once. Conversely the normalisation process is rather simple requiring just  a multiplication with a 6-cycle latency followed by a bit shift.

The heart of the processor is a 32-bit up/down accumulator. This accumulator is reset at the start of each pixel, not to zero but to a "digital bias". This adds a small positive offset (1024ADU is convenient) that ensures that the accumulator never goes negative, as it could in the case of a CCD bias image. After reset, weighted reference samples are accumulated until the end of the pedestal is reached. The polarity is then flipped and the weighted *signal* samples accumulated.

The final processed pixel value appears at the output 20 ADC clock cycles after the end of the signal pedestal. The data is 18-bit wide although no investigation was done on the validity of the two bits that follow the implicit binary point.

The algorithm was initially simulated with Python to ensure there were no significant rounding errors or any saturation effects regardless of the choice of coefficients and input signal amplitude. This simulation  showed the need to use an output scaler of 32-bit resolution.

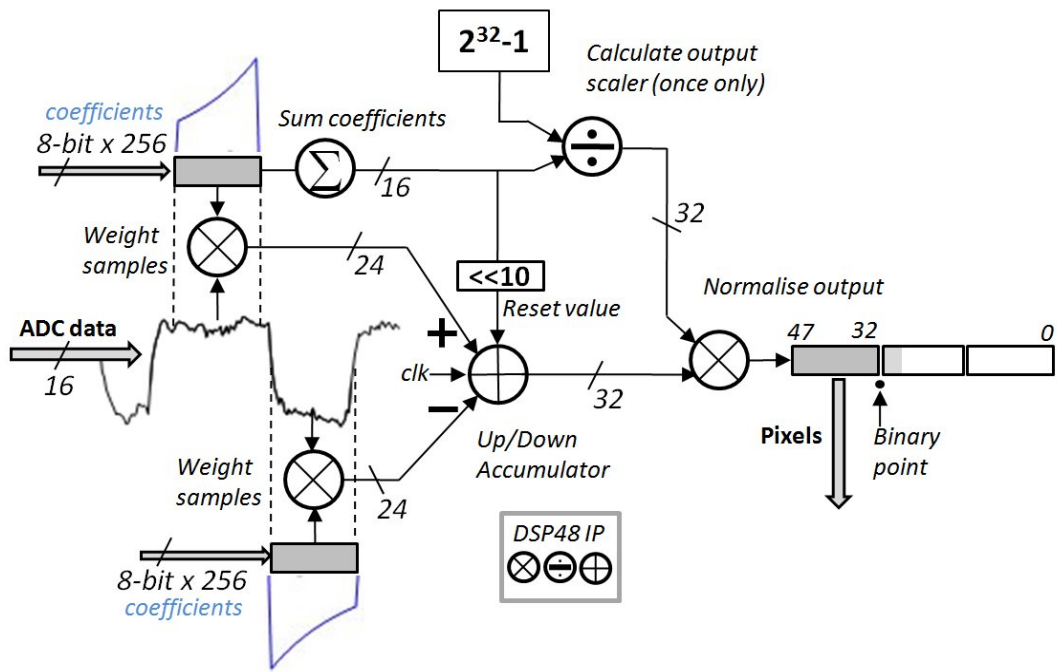

**Fig. 5.** The DCDS Algorithm as  implemented in the VHDL hardware

## 4 FPGA implementation

### 4.1 The Opal Kelly FrontPanel

The FrontPanel system comes included with the Opal Kelly FPGA module and consists of Windows drivers and a Python SDK for the host and special IP to run on the FPGA. The IP allows the user to instantiate virtual parallel ports and high speed data pipes within the FPGA. Data can be moved to and fro between host and module in a very simple fashion. The transfer speed was limited by the USB2 interface to about 38MB/s, sufficient to continuously stream oscilloscope-mode data if the ADC was slowed to 10MHz.

### 4.2 Development process

The various IP elements that made up the DCDS processor were first behaviorally tested using the integrated Vivado test bench. The implemented designs were then verified using FrontPanel to supply test data and read back results to the host PC. The final stage was to supply a synthetic CCD video waveform to the preamp, provided by a DAC. This DAC could simulate accurately the video waveform from a CCD in four modes of operation:

-A bias frame for noise measurement
-A ramped exposure for linearity and dynamic range tests
-A flat field mode with all pixels of constant level
-A bias frame with every 16th column set to saturation to verify the impulse response.

The synthetic DAC control IP and physical hardware removed the need for an operational CCD and an optical bench; the DCDS hardware could be substantially tested using just the development board, a PC and a 5V power supply.

### 4.3 Resource usage

The post implementation floor plan is shown below. It fits comfortably into the chosen Artix device. A smaller device such as the xc7a15tftg256 would still have accommodated at least two but maybe as many as four DCDS processors in a 17 x 17mm footprint.

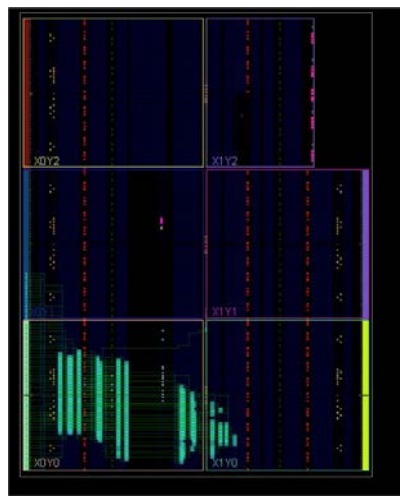

DSP48E1   x5  (4.2%)
Block RAM   <1tile  (<1%)
Slice LUTs x995  (3%)
Slice Registers x3547 (5.4%)

**Fig. 6.** Resource utilization by a single channel DCDS processor on an Artix xc7a50tifgg484-1L

## 4.4 Port description

The block diagram representation of the DCDS processor produced by the Vivado design suite is shown below.

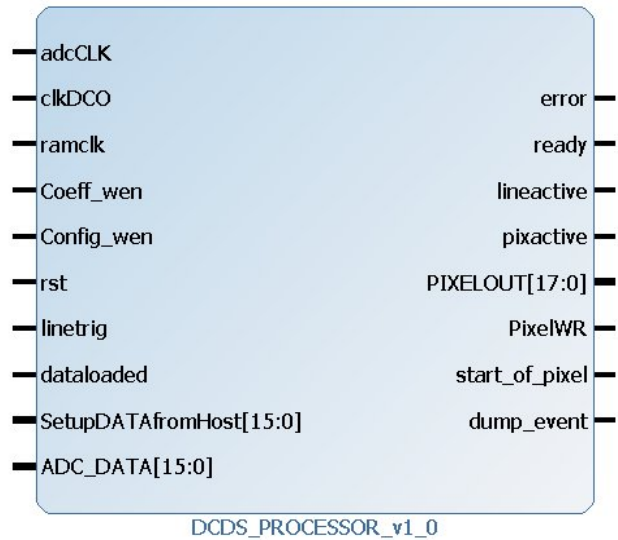

**Fig.7.** Top level block diagram of the DCDS processor

**adcCLK**: 20MHz conversion clock used by ADC (in).

**clkDCO**: echoed clock received from AD9269 used in project (in)

**ramclk**: clock used to load coefficient and configuration registers (in).

**Coeff_wen**: enable writes to coefficient RAM (in)

**Config_wen**: enable writes to configuration registers (in)

**rst**: global reset (in).

**linetrig**: Trigger input to start a readout of 1 line of image (in).

**dataloaded**: Trigger from host to signify register/RAM data has been fully sent (in)

**SetupDATAfromHost**: Data bus from host for registers/coefficient RAM

**ADC_DATA**: Parallel Data from ADC

**error**: flag that configuration data is inconsistent (out)

**ready**: the DCDS processor is configured and ready for a line trigger (out)

**lineactive**: a line of the image is currently being read (out)

**pixactive**: a pixel read is currently active (out)

**PIXELOUT**: Processed pixel parallel output

**PixelWR**: Pixel data strobe (out)

**start_of_pixel:** trigger to external CCD serial clock generation circuitry (out)

**dump_event**: trigger to external CCD serial clock generation circuitry (out)

## 4.5 Internal registers

A typical CCD video waveform covering approximately one pixel period is shown in figure 8. overlaid with the key parameters that define its timing.

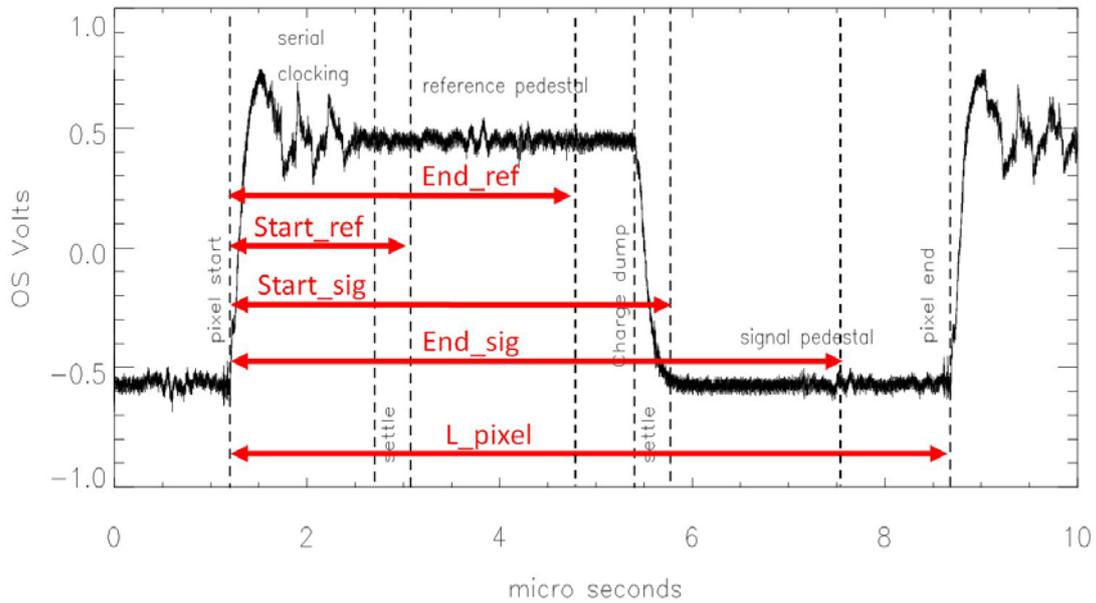

**Fig.8.** Description of pixel waveform

The purpose of the DCDS processor is to faithfully measure the size of the step in this waveform that appears the instant the photo-charge is "dumped" onto the output node of the CCD. This dump event is preceded by a reference pedestal and followed by a signal pedestal. It is up to the user to specify which parts of the waveform are measured by the DCDS and this is done by way of the configuration registers.

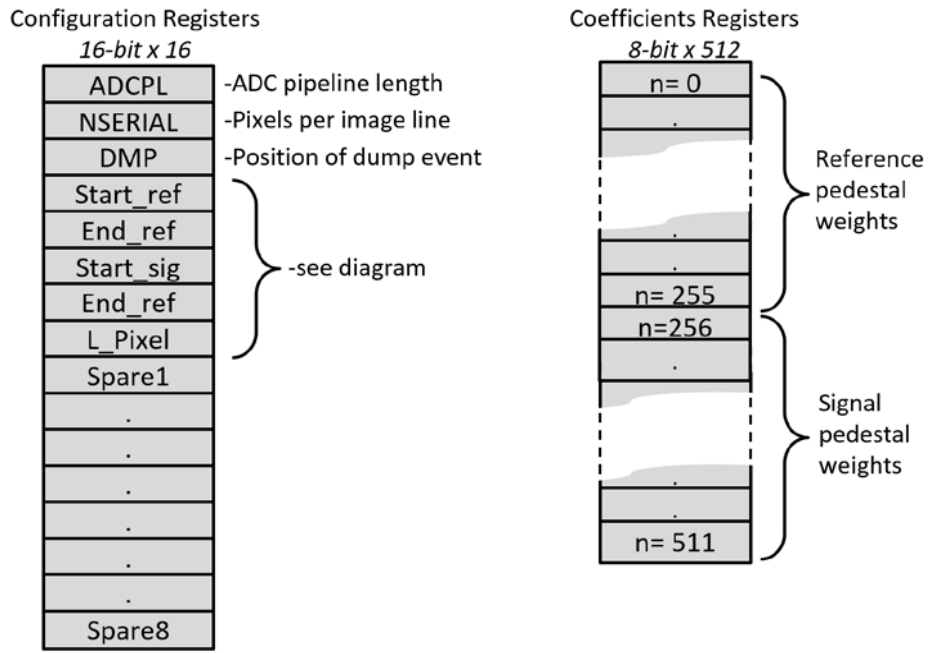

**Fig.9.** Description of DCDS processor registers

A second bank of registers defines the weighting coefficients applied to the samples in each pedestal. These pedestals can be up to 256 ADC samples long. The sum of the weights in each pedestal is typically the same. Combinatorial logic checks that the pedestals are of equal length and non-overlapping. It also checks that the final signal pedestal sample does not occur within 5 cycles of the end of the pixel, this being the time it takes for the pixel sequencer FSM to return to its triggerable state.

## 5      Performance tests

The five main ways that a DCDS system could misbehave were identified as:

- Sudden jumps in pixel value in a graded image due to rounding effects
- Saturation effects limiting dynamic range to below 16-bits
- Persistence effects resulting in one pixel leaking into its neighbour
- Excess noise whereby rounding errors in the DCDS algorithm exceed the analogue noise.
- Framing errors between desired and actual active-pedestal regions.

### 5.1     Linearity tests

Here a series of ramped images were taken using the DAC waveform generator, one of which is shown in figure 10. Since the DAC only had an 8-bit resolution the ramp had visible steps. A horizontal cut through this image then proved that the system could reach the full 16-bit dynamic range. Various sample weighting schemes were used to generate further images. It was established that setting all weights to 1 or setting all weights to 255 generated the same image. As an additional check, a sloped weight profile with the first sample in each pedestal=0 and the last =255 was used.

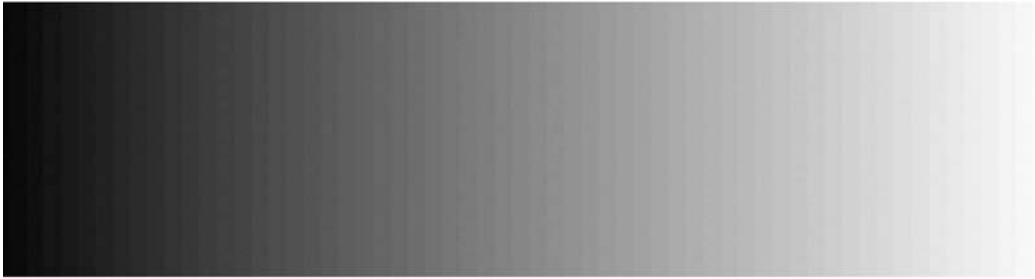

**Fig.10.** A ramped image used to test linearity and gain stability

The 8-bit DAC had insufficient resolution to test for CDS rounding errors so an alternative technique was used. One nice thing about a DCDS system is that it can be run in 'oscilloscope' mode by directing the raw ADC samples directly to the output FIFO normally used for the processed pixel values. This can be very useful for analogue optimisation of the preamp and, with subsequent FFT analysis, for the identification of extraneous noise sources. The oscilloscope mode was used to grab a sequence of ADC samples corresponding to 1 line of the DAC-generated ramp image. The lower-most bit of the sample data was discarded and replaced with a flag bit that signified which samples were being acted upon by the CDS in its normal course of operation i.e. the active samples within each pedestal. This raw ADC data, an example of which can be seen in figure 13, was then analysed using a CDS algorithm implemented in IDL on the host PC. A further identical ramp image was then taken but in normal CDS mode and the calculated pixel values compared to those produced by the IDL program. The difference between the two results was then compared to test the computational accuracy of the DCDS algorithm as implemented in the FPGA hardware. The results are shown in figure 11. Errors were within +/-5ADU up to 50kADU.  Beyond this the divergence is thought to be due to analogue  saturation in the preamplifier.

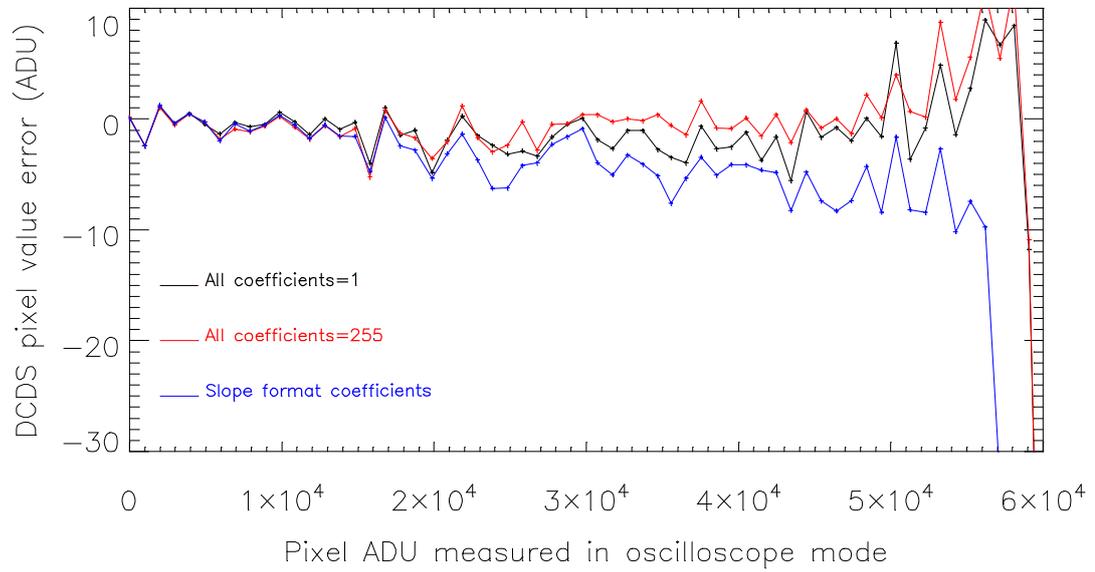

**Fig. 11.** Result of the linearity tests

### 5.2 Impulse response

The DAC-based synthetic CCD waveform generator was programmed to give a single line of increasing brightness across the image every 16 columns (see figure 12) to check for persistence effects in the CDS. None was seen confirming that the sample accumulator was being reset after each pixel.

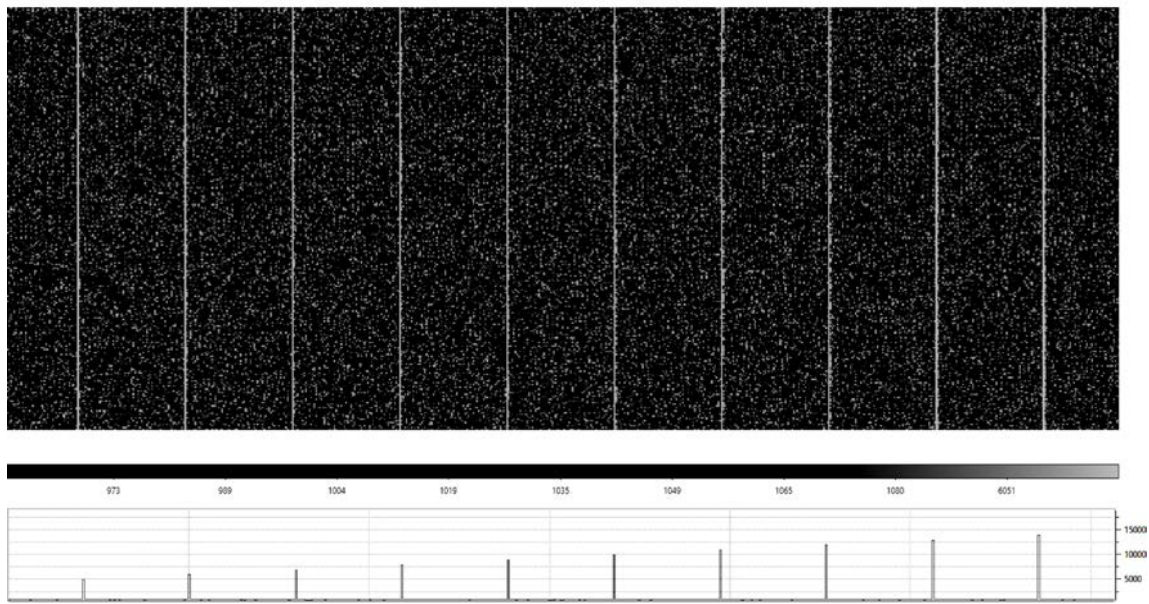

**Fig. 12.** An image with bright columns used to verify the impulse response of the DCDS system.

### 5.3 Framing error detection

Figure 13. shows a 100μs section of raw ADC data taken in 'oscilloscope mode' and covering 10 pixels. The active regions of the waveform, which are highlighted in red, were flagged by the CDS and encoded into the lower bit of the data. It is clear that they have been correctly placed. Note that the first part of this waveform is contaminated by the clock feed-through from the baseline restoration circuitry which has correctly placed the reference pedestals close to the top of the ADC range.

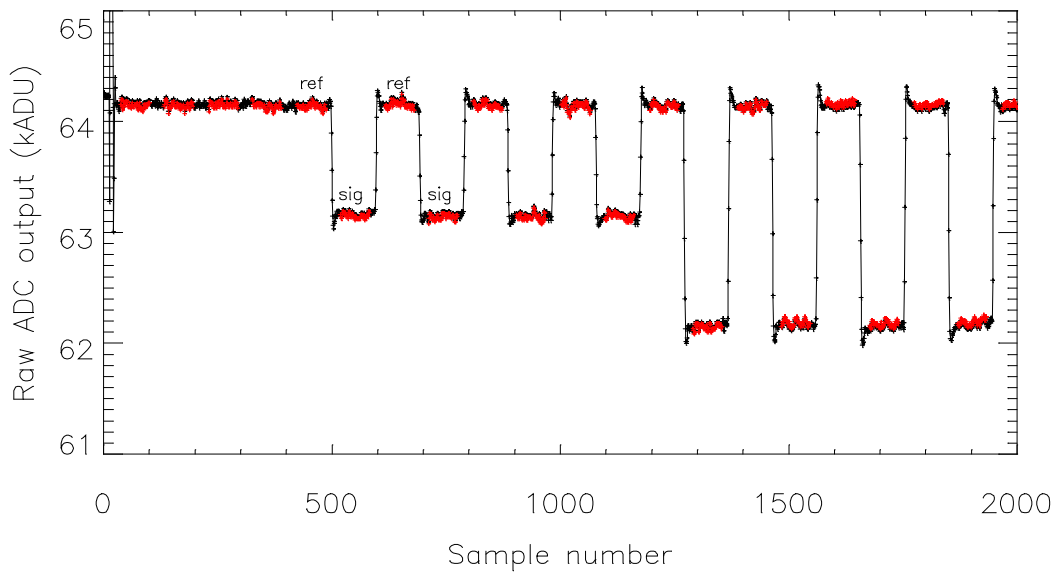

**Fig. 13.** Raw 20MHz output of the ADC in response to part of the ramp-image readout (see figure 10). The lower bit of the ADC data was used to flag which samples were used by the DCDS processor. These are highlighted in red.

### 5.4 Noise

The whole point of the DCDS is to reduce noise so it was important to demonstrate that the overall system noise would be dominated by the CCD connected to it. To test this, the input of the preamplifier was connected to ground via a 380Ω resistor to simulate the typical source impedance of a CCD. A series of bias frames was then taken with all samples weights set to 255 and the standard deviation in the image plotted as a function of the number of samples in each pedestal. This was done for two preamplifier gains: x2.4 corresponding to 1.7 $e^-$/ADU and x4.9 corresponding to 0.8$e^-$/ADU system gain (when used with a CCD231). The result is shown in figure 14. The noise in both cases bottomed out at <1ADU, although in the graph the result is shown in units of μV RMS referenced to the front end. The x-axis of this graph is expressed both in units of samples-per-pedestal and in pixel rate. The pixel rate here may seem rather slow given the corresponding values for the pedestal widths but the test system was severely limited by the rather slow settling time of the DAC used to generate the CCD waveform. As can be seen, sub-electron CDS noise was achieved with a preamplifier differential gain of x4.9 and for pedestal widths in excess of 2.5μs.

### 5.5 Logic speed

For simplicity the DCDS ran from a single 20MHz clock; that used by the ADC. This is quite slow for an Artix-7 device and no timing problems were encountered. As an experiment the clock frequency was increased to see at what point the hardware failed. Even at 60MHz a sensible slope image was obtained although the ADC was by this point quite non-linear. At 80MHz the system failed completely. The AD9269 is available in 20,40 and 65MHz pin-compatible variants so it would be interesting to measure the DCDS behaviour with these alternative devices.

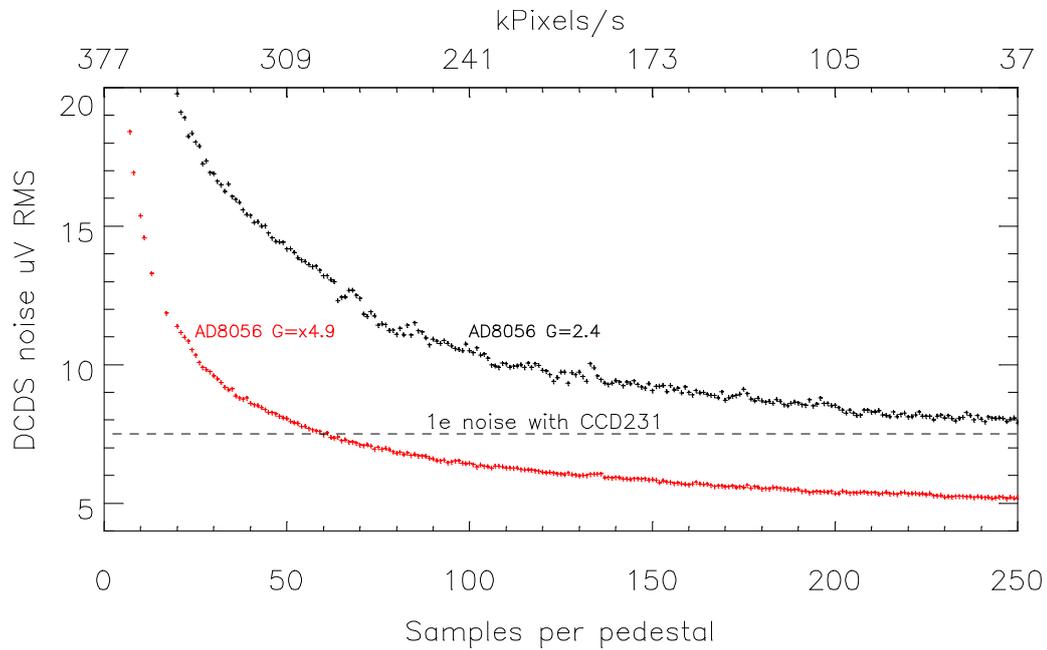

**Fig. 24.** The front-end referenced noise of the DCDS system with grounded input. ADC frequency=20Msps.

## 6      Conclusions and further work

The DCDS system worked well. It remains to test it with a live CCD. The limited differential input range of the ADC meant that relatively high preamplifier gains were needed to overcome noise in the ADC and its differential driver IC. This limited the dynamic range to around 55ke$^-$. One solution would be to use an ADC with a wider input range (from the AD Pulsar family), alternatively the ADC driver IC could be equipped with a simple switched attenuator to reduce the gain by a factor of 2 to 3 for exposures requiring the full-well range of the detector.

In the future the DCDS hardware could be implemented on a Xilinx Zynq processor with a data acquisition system running under Linux on one of the embedded ARMs. A Linux driver would then take the place of the FrontPanel interface. This 'system on a chip' approach could result in a very compact CCD controller but will require considerable expertise in the driver development. Another avenue would be to build a compact video front-end with a fully digital output that could be placed either within or just outside the CCD cryostat and communicate with an existing DAS host system over potentially very long cables with no degradation in read noise. Figure 15. shows the general idea.

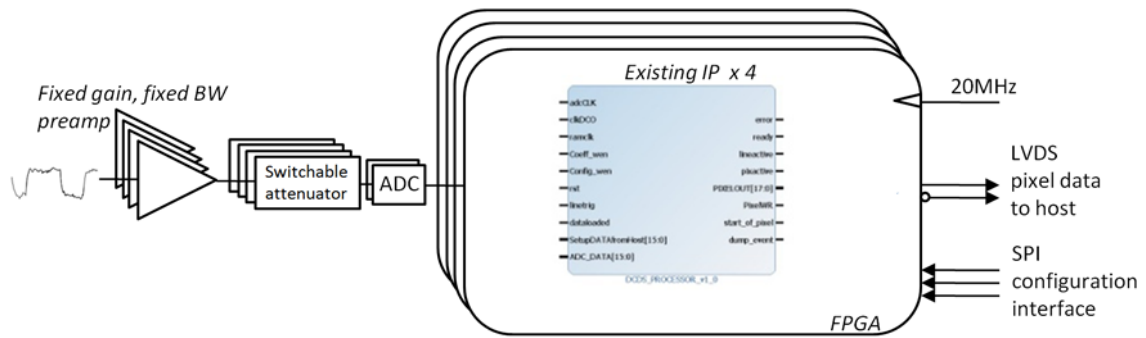

**Fig. 35.** Idea for an extended 4-channel system. This could fit within the footprint of a 4k x 4k CCD.